\documentclass[10pt]{article}
\usepackage{epsf}
\newcommand{\beq}{\begin{equation}}
\newcommand{\eeq}{\end{equation}}

\begin{document}

\title{Dark Matter Accumulation near the Earth for the Long Range Forces
case }

\author{M.S.~Onegin\footnote{E-mail: onegin@thd.pnpi.spb.ru}, A.P.~Serebrov,  O.M.~Zherebtsov\\
Petersburg Nuclear Physics Institute, Gatchina, St.~Petersburg
188300, Russia }

\maketitle

\begin{abstract}
The accumulation of dark matter near the Earth is considered. We
analyze the case of long range interaction forces. Additional
density of the dark matter at the Earth's surface is calculated.
We show that this density exceeds the mean density of the dark
matter in our galaxy by more then $10^5$ times for some values of
dark matter particle mass. Accumulation of WIMP's near the Earth
by the same mechanism is also analyzed.
\end{abstract}

\section{Introduction}

The dark matter (DM) in our Galaxy interacts with the Earth. The
mean velocity of DM particles in the Galaxy is about 220 km/s.
When such a particle is elastically scattered on the nuclei of the
Earth it loses its energy and could be gravitationally absorbed by
the Earth. As the escape velocity of the earth is equal to 11.2
km/s the DM particle should be slowed down to lower velocity to be
captured. We are considering the following form of the potential
of DM and ordinary particles

\beq V(r)=G \frac{m_x m}{r}(1+\alpha \frac{e^{-\lambda r})}{r}
\eeq

Here $m_x$ - is the mass of the DM particle, $m$ - mass of the
ordinary particle or body, $\alpha$ - parameter which  is
characterizing the relative strength of the gravitational force
and the long range forth (LRF), $\lambda$ - is the range of the LRF
interaction. In the previous parer~\cite{Serebrov} the possibility
of accumulation of DM with LRF  in the Sun system was assumed and
in this paper we also fix the parameter $\alpha$ to be equal to
$10^{26}$. The DM particle mass $m_x$ and parameter $\lambda$ will
vary.

After the DM particle was gravitationally absorbed it can stay at
the stationary orbit around the Earth until it is captured by
the Earth. Staying inside the Earth the DM particle will be
thermalized and get velocity distribution determined by the
temperature of the Earth in the core. This temperature could be as
high as 6000 K and the DM particle has noticeable probability to
fly out of the Earth and in some cases even to escape the Earth if its
velocity is higher than 11.2 km/s.

In this paper we will calculate capture rate of the DM particles
by the Earth, and also the time life of them in the Earth
$t_{Earth}$ and inside the Earth gravitation $t_{life}$. Using
these quantities we can calculate the additional density of DM
particles at the Earth surface. As it follows in some cases this
additional density will considerably exceed the DM mean density in
our Galaxy.

\section{Earth model}
\begin{figure}[!b]
\centerline{ \epsfxsize=12cm\epsfbox{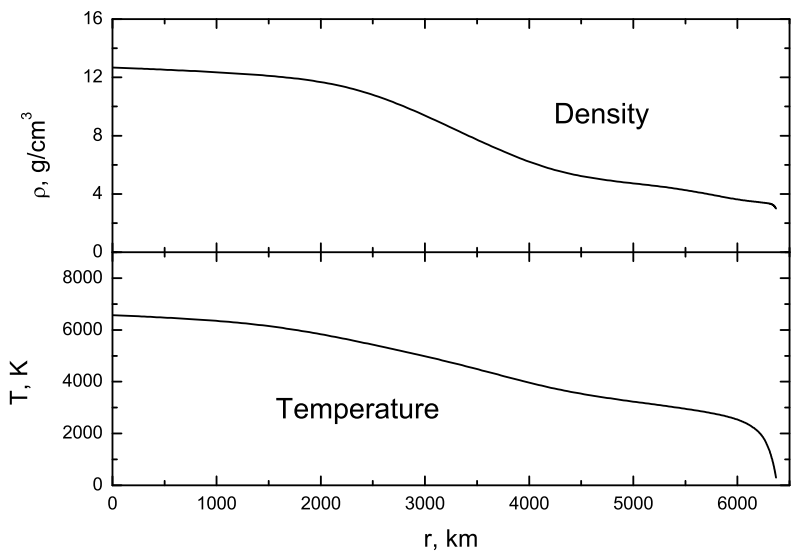}} \caption{Radial
distribution of the Earth density and temperature} \label{EP}
\end{figure}

The main properties of the Earth needed to describe the
propagation of the DM in it are radial  dependencies of the Earth
density and  the temperature, as well as the Earth elemental
composition. Radial dependence on the Earth density and
temperature we used are presented in Figure~\ref{EP}. We
considered the Earth composition in the mantle and in the
core to be different. Elemental composition of the Earth's mantle and core is
presented in Table~\ref{EC}~\cite{Lundberg}. The depth of the
mantle is equal to 2900 km. In our calculation we assume that
the mantle consists of the single element with average atomic
mass equal to 23.5 and the core consists of iron only. We
also divide the Earth volume into concentric layers with constant
density and temperature obtained by averaging the
real density and temperature distribution over the layer volume. The
number of layers we considered was equal to 20.

\begin{table}[hb!]
\caption{The composition of the Earth's core and mantle}
\begin{center}
\begin{tabular}{|l|c|c|c|}
\hline \hline & Atomic & \multicolumn{2}{|c|}{Mass fraction} \\
\cline{3-4}
 Element & number & Core & Mantle \\
\hline
Oxygen, O & 16 & 0.0 & 0.440 \\
Silicon, Si & 28 &0.06 & 0.210 \\
Magnesium, Mg & 24 & 0.0 & 0.228 \\
Iron, Fe & 56 & 0.855 & 0.0626 \\
Calcium, Ca & 40 & 0.0 & 0.0253 \\
Phosphor, P & 30 & 0.002 & 0.00009 \\
Sodium, Na & 23 & 0.0 & 0.0027 \\
Sulfur, S & 32 & 0.019 & 0.00025 \\
Nickel, Ni & 59 & 0.052 & 0.00196 \\
Aluminum, Al & 27 & 0.0 & 0.0235 \\
Chromium, Cr & 52 & 0.009 & 0.0026 \\
\hline \hline
\end{tabular}
\end{center}
\label{EC}
\end{table}

\section{DM propagation in the Earth}
We consider separately the influence of the gravitation and LRF on
DM particle propagation in the Earth. To simplify simulation of DM
propagation in the gravitational potential of the Earth we
approximate potential to be constant in the layers into which we divide the
Earth. As in region with constant potential the gravitational
force is zero, the DM particle propagates freely inside the layers.
The trajectory of particle breaks only at the border of layers.
The particle can be refracted or reflected from the border
surface. The law of refraction has the following form \beq
\left\{ \begin{array}{ll} v_1^2/2+\phi_1=\frac{v_2^2}{2}+\phi_2
\\ v_1 \sin \theta_1 = v_2 \sin \theta_2 \end{array} \right.
\eeq where $\phi_{1,2}$ are the gravitational potentials in
regions 1 and 2, $v_{1,2}$ - velocity of particle in these
regions, $\theta_{1,2}$ are the angles between the particle
trajectory and the normal to the  surface at the point where
trajectory intersects the surface. In the case when equation system
(2) has no solution the particle reflects mirrorly from the
surface.

\begin{figure}
\centerline{ \epsfxsize=12cm\epsfbox{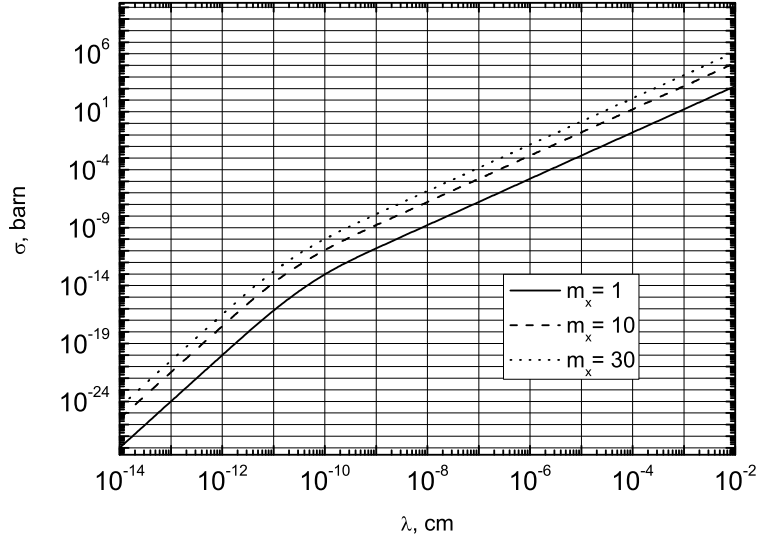}} \caption{Total
scattering cross section of DM particle for different $\lambda$}
\label{CS}
\end{figure}

Scattering on Yukawa potential of (1) can be treated in Born
approximation for interaction length $\lambda < 3\times 10^{-2}$
cm. Scattering amplitude in the c.m. system has the following form
\beq f(q)=\frac{2 \mu}{\hbar^2} a \frac{\lambda^2}{(\lambda
q)^2+1}, \eeq where $\mu=\frac{m_x m}{m_x+m}$ is a reduced mass of
scattering particles, $q$ is wave vector transfer, and parameter
$a=-\alpha G m_x m$. Total cross section of scattering for
different $\lambda$ and $m_x$ for $m=1$ GeV is presented in
Figure~\ref{CS}. As is seen from the figure total cross section
strongly depends on the interaction length and DM mass. For
$\lambda > 10^{-6}$ cm interaction length becomes rather short and
the simulation of DM propagation becomes very time consuming.
Scattering angle in the c.m. system can be simulated according to
the following formula \beq \cos \theta = \frac{1}{b}
[b+1-\frac{1+2 b}{1+2 b \xi}], \eeq where $\xi$ is a uniform
distributed in the interval [0,1] random number. Parameter
$b=2\mu^2v^2\lambda^2/\hbar^2$  determines the asymmetry of
scattering. The velocity distribution of Earth particles is taken
to be Maxwellian with the temperature of the layer where
scattering occurs.

\section{Capture rate of DM particles with the Earth}
We take the velocity distribution of the Galaxy DM particles to be
Maxwellian~\cite{Damour}, \beq f_0(v)dv=n_x
\frac{4}{\pi^{1/2}}\frac{v^2}{v_0^3}e^{-v^2/v_0^2}dv \eeq where
the parameter $v_0=220$ km/s. We have taken the energy density of
DM $\rho_x=n_x m_x=0.3$ Gev/cm$^3$. As the Sun with the Earth is
moving in the Galaxy coordinate system with the velocity $v_S=220$
km/s, the velocity distribution of the DM particles, as we see it
from the Earth has the following form: \beq f(v)=n_x
\frac{1}{\pi^{1/2}} \frac{v}{v_0
v_S}[e^{-(v-v_S)^2/v_0^2}-e^{-(v+v_S)^2/v_0^2}]. \eeq Velocity of
DM particle increases when it approaches the Earth surface from
the infinity according to the following equation: $v'^2=v^2+\frac{2 G
M_E}{R_E}$. We consider the flux of DM particles to be isotropical
neglecting its asymmetry due to the motion of the Sun.

After the DM particle falls on the Earth its propagation is
considered according to the previous paragraph. During the elastic
scattering on the earth nuclei the DM particle loses its energy. We
consider the DM particle to be captured by the Earth if it is
reflected several times from the surfaces of layers into which the
Earth volume was divided. Knowing the initial flux of DM and
calculating in this manner the probability of capturing we have
calculated the capture rate of DM by the Earth.

\subsection{Model testing}
\begin{figure}[!h]
\centerline{ \epsfxsize=12cm\epsfbox{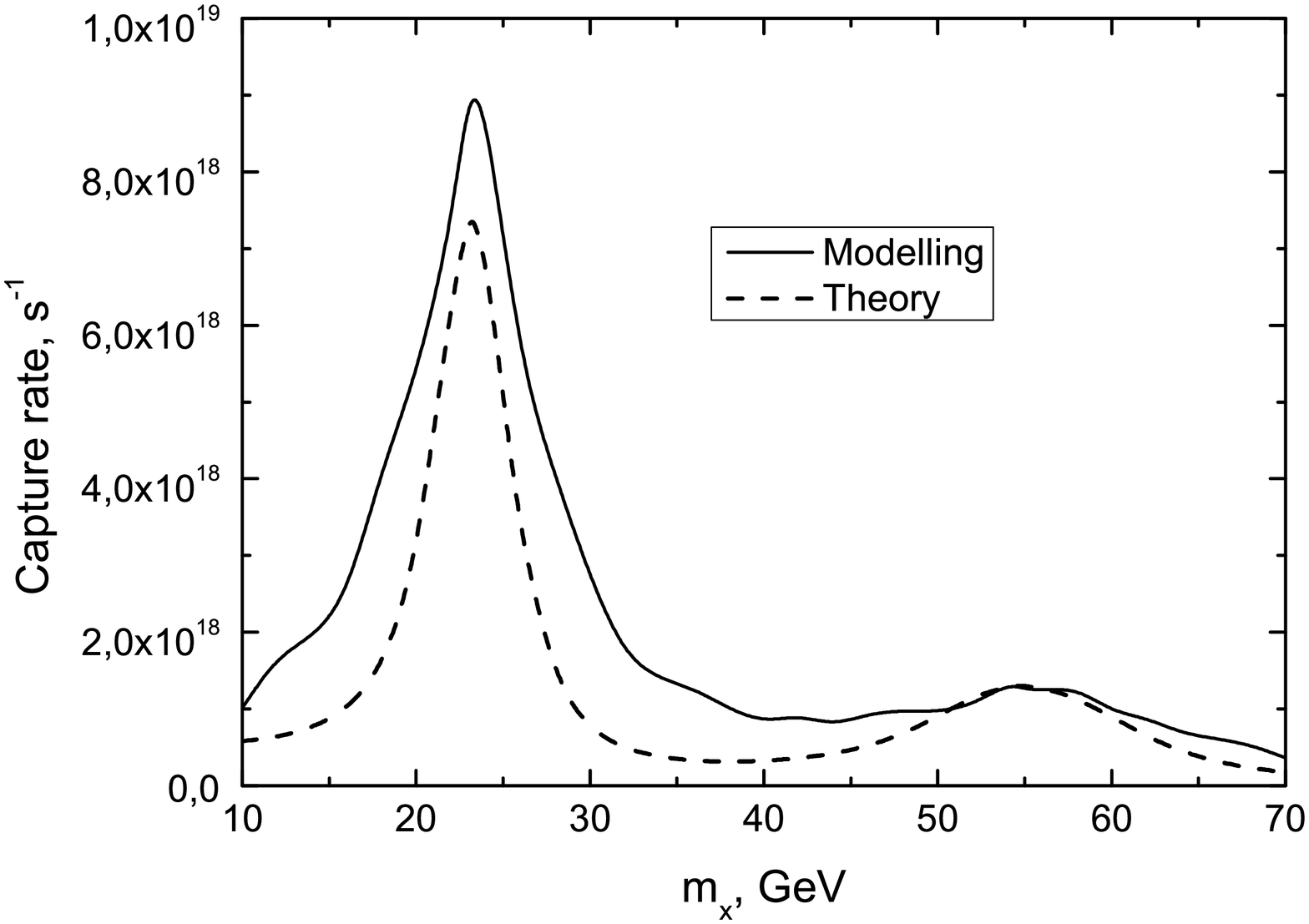}} \caption{ Capture
rate of WIMP's of different mass by the Earth} \label{Caprate1}
\end{figure}

To test our model we have tried to calculate the test case of
capture of weak interaction massive particles (WIMP) by the Earth.
We suppose that these particles have fixed scattering cross
section $\sigma_0$ with the nuclei of the Earth, and that this
scattering is elastic and isotropic. We took this cross section to
be equal to $\sigma=10^{-34}$ cm$^2$. In paper~\cite{Gould} the
capture rate of WIMP's by the Earth was theoretically calculated.
In this test calculation we neglect the motion of the Solar system
relative to the Galaxy coordinate system when the analytical
formulas for the capture rate are the most transparent. When the
Sun motion is taken into account the caption rate changes only by
a factor not higher than 2.5. Also analytical formula doesn't take
into account non zero temperature in the Earth. As was shown
in~\cite{Gould} the influence of non zero Earth temperature on the
capture rate is not crucial. Every element of the Earth gives its
own contribution to the caption rate. Consider element of mass $m$
with number density n. Let its mass fraction in the Earth be $f$.
It is convenient to introduce the following notations: \beq
\begin{array}{cl}
\mu=\frac{m_x}{m},\;\;\; \mu_\pm=\frac{\mu \pm 1}{2}\\
A=\frac{u^2}{v_0^2}\frac{\mu}{\mu_-^2} \end{array}, \eeq where $u$
- is escape velocity of DM particle from the considered point inside
the Earth. Let also $\hat\phi$ be the gravitational potential in
the Earth relative to this potential at the Earth surface. The
value of $\hat\phi$ is about 1.6 in the Earth center. We denote
with $v_{esc}$ the escape velocity of particle at the Earth
surface. This value is about 11.2 km/s. Then the caption rate of
the DM by the considered element can be calculated
as~\cite{Gould}:

\beq C=[(\frac{8}{3\pi})^{1/2} \sigma n_x \bar
v][\frac{M_B}{m}][\frac{v_{esc}^2}{v_0^2}<\hat\phi>][\langle
\frac{\hat\phi}{<\hat\phi>}(1-\frac{1-e^{-A^2}}{A^2})\rangle],
\eeq where $\bar v=\sqrt{\frac{3}{2}}v_0$, $M_B$ is the mass of
the Earth and angle brackets indicate averaging over the mass of
the Earth. The total capture rate for all elements can be
calculated as $\sum_f f C_f$ where $C_f$ is the capture rate
calculated according to equation (8). DM mass dependence on the
capture rate for individual element has a maximum at $m_x=m$.
Theoretically the calculated capture rate of WIMP's according to this
formulas for our model of the Earth is presented in
Figure~\ref{Caprate1}. Here the result of our Monte Carlo
simulation is also presented in comparison. We see that mass
dependence on the theoretically calculated capture rate has two
maxima according to two average elements we have considered: one
with mass $m=23.5$ GeV in the mantle of the Earth and the second
at $m=56$ GeV in the core. Monte Carlo results qualitatively
agree with a theoretical prediction. Its mass dependence also has
two maxima but the first maximum is about 20\% smaller than the
theoretical prediction and the width of the theoretical peaks is
about two times smaller than in our model. Probably this
discrepancy between curve forms can be a consequence of neglecting
the temperature of the Earth in the theoretical formulae.

\subsection{Capture rate of long range DM by the Earth}

\begin{figure}[!hbp]
\centerline{ \epsfxsize=12cm\epsfbox{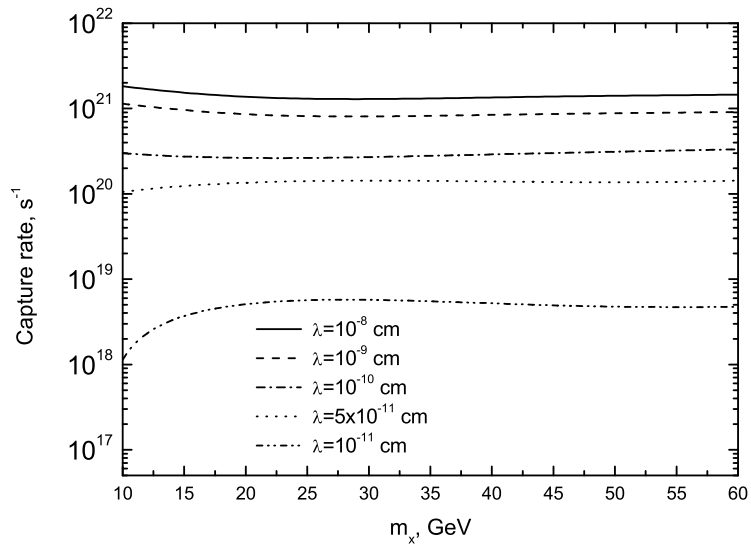}} \caption{ Capture
rate of DM by the Earth for different $\lambda$} \label{Caprate2}
\end{figure}

We have calculated the mass dependence on the capture rate for a
long range interaction of DM with ordinary matter for different
values of $\lambda$. Results are presented in
Figure~\ref{Caprate2}. For a long range interaction DM the mass
dependence of capture rate is smooth without maxima. This is due
to the fact that scattering cross section for such matter is high
and DM particle has a lot of interaction in the Earth with small
transfer of energy during each interaction. Also the absolute
value of capture rate for $\lambda > 10^{-10}$ cm is about two
orders of magnitude larger than for WIMP's with cross section
$\sigma=10^{-34}$ cm$^2$. Capture rate increases with parameter
$\lambda$ and is even higher for larger $\lambda$ than shown in
Figure~\ref{Caprate2}.

\section{DM evaporation}

\begin{figure}[!hbp]
\centerline{ \epsfxsize=12cm\epsfbox{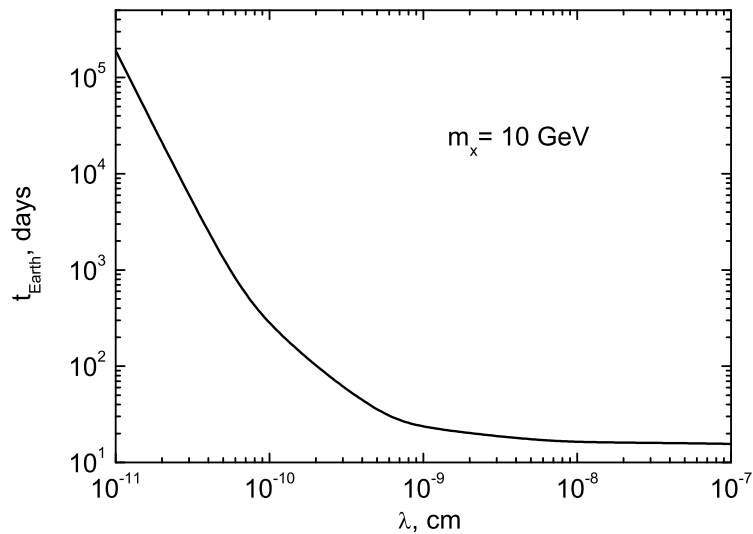}} \caption{Time life
in the Earth for different $\lambda$} \label{timeE}
\end{figure}

Captured DM thermolizes in the Earth so that its position-velocity
distribution follows Maxwell-Boltzmann law: \beq
f_{th}(v,r)=\frac{n_0}{V_1}\frac{4}{\pi^{1/2}}(\frac{m_x}{2k_BT_x})^{3/2}
v^2e^{-m_x v^2/2k_BT_x}e^{-m_x \Phi(r)/k_BT_x}, \eeq where $n_0$ -
is the number of captured particles, $T_x$ - temperature of DM
distribution, $\Phi(r)$ is the Earth gravitational potential, and

\beq V_1=\int_{0}^{R_E} 4\pi r^2 e^{-m_x\Phi(r)/k_BT_x}dr \eeq

is the effective volume of the Earth and $R_E$ is the Earth
radius. Particles captured by the gravitational potential of the Earth
can fly out of the Earth, move along the elliptic orbit
around the Earth and then again return into the Earth. Sometimes if
the velocity of the particle is larger than the escape velocity of
the Earth the particle can fly out of the Earth. The latter process
determines the total timelife of the captured DM particle. The
rate at which DM particle fly out of the Earth is determined by
the mass of DM particle, difference of gravitational potential at
the Earth surface compared with the center of the Earth and by the
cross section with which DM particle interacted in the Earth. In
Figure~\ref{timeE} time life of DM particle in the Earth is
presented for particle of mass equal to 10 GeV and for interaction
of different length $\lambda$.

\begin{figure}[!hbp]
\centerline{ \epsfxsize=12cm\epsfbox{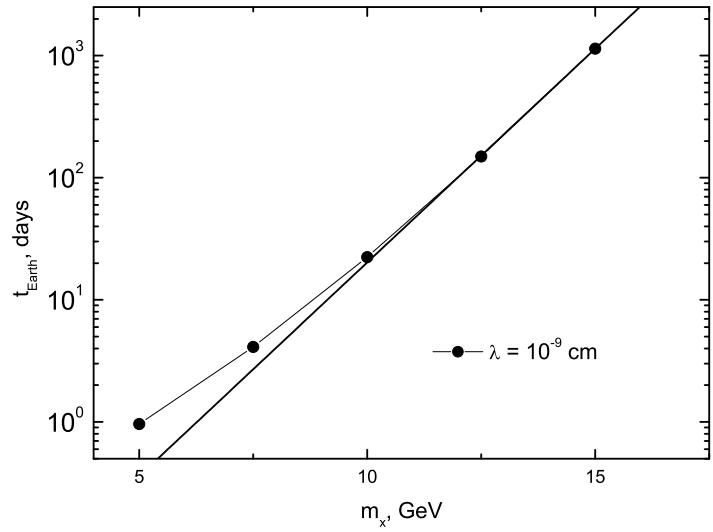}} \caption{Time
life in the Earth for different $m_x$} \label{timeE2}
\end{figure}

Time life of DM particle in the Earth depends on the mass of DM
particle $m_x$ according to the Boltzmann distribution: \beq
t_{Earth}^{-1} \sim e^{- m_x \Delta \Phi /k_B T_x}, \eeq where
$\Delta \Phi$ is the difference of the gravitational potential
between the center and the surface of the Earth. Calculated time
life of DM particle in the Earth for different $m_x$ is compared
with this distribution in Figure~\ref{timeE2}. Calculated time
life in Figure~\ref{timeE2} for small $m_x$ diverges from simple
law (4) because the particles flying out of the Earth have nonzero
kinetic energies.

\begin{figure}[!hbp]
\centerline{ \epsfxsize=12cm\epsfbox{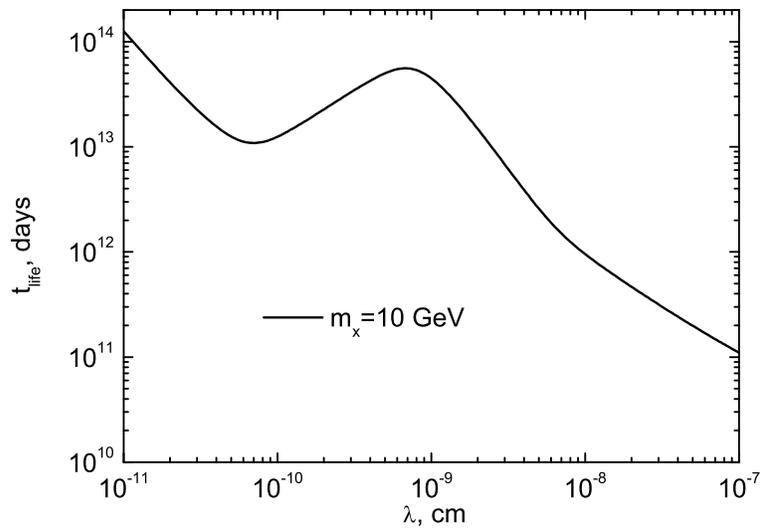}} \caption{Total
time life of DM particle before escaping for different $\lambda$}
\label{timelife}
\end{figure}

Total time life of the captured DM particle in the Earth
gravitation is presented in Figure~\ref{timelife} for particle
with mass $m_x=10$ GeV for different $\lambda$. It has a peak for
interaction length $\lambda$ equal to $10^{-9}$. Total time life
in the Earth gravitation is determined by the time life in the
Earth $t_{Earth}$ which is divided by the probability of escaping
the Earth. The latter probability depends on the form of the
velocity spectrum of the particles which leave the Earth. Computer
simulation shows that this spectrum for $\lambda > 10^{-9}$ can be
described by simple Gaussian with nearly constant dispersion. On
the other hand, the mean velocity of DM particle flying out of the
Earth slightly rises with $\lambda$ here so that probability of
escaping the Earth gravitation increases and the total time life
in Fig.~\ref{timelife} falls for large $\lambda$.

\begin{figure}[!hbp]
\centerline{ \epsfxsize=12cm\epsfbox{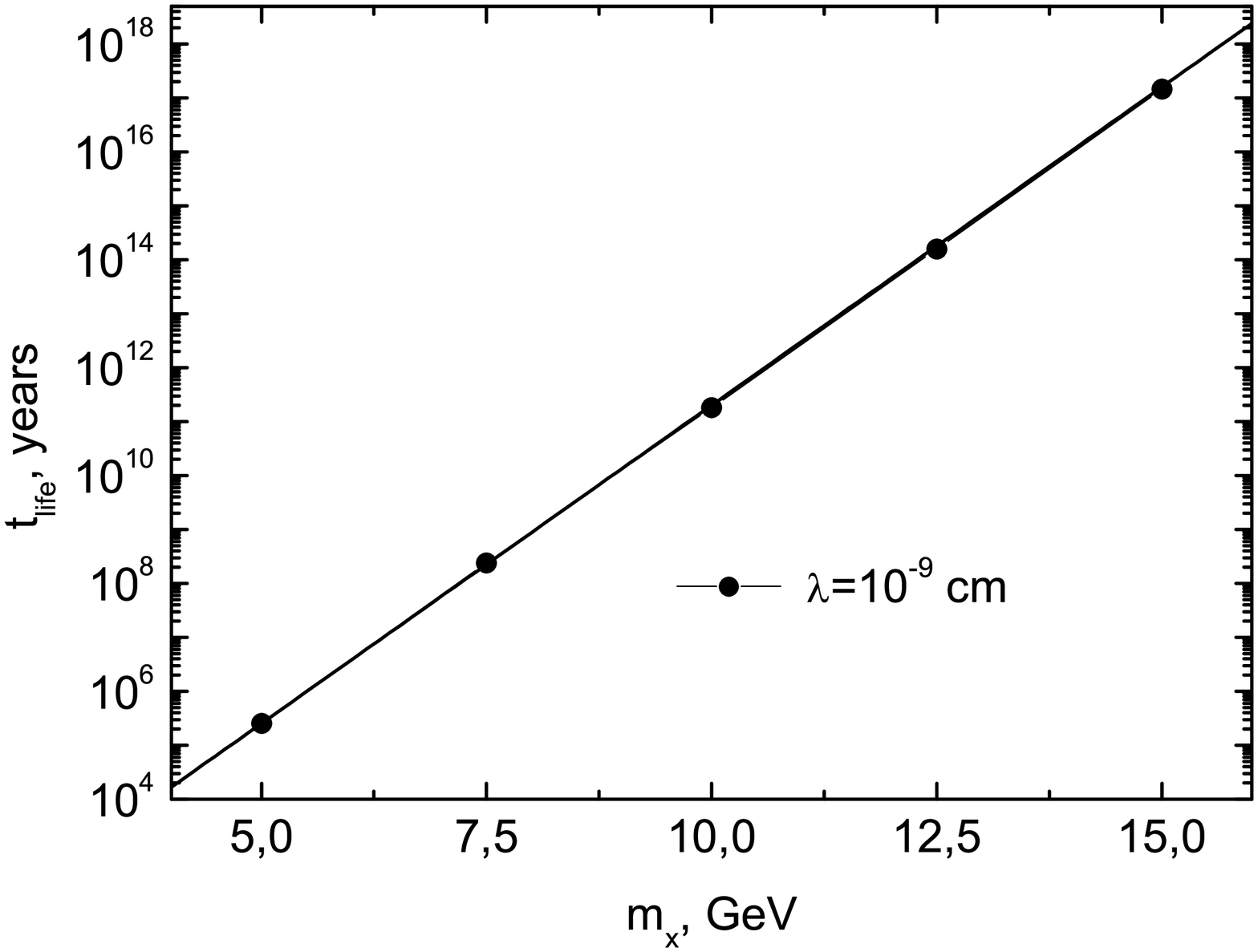}} \caption{Total
time life of DM particle before escaping for different masses of
DM particles} \label{timelife2}
\end{figure}

Dependence of total time life of DM particle on $m_x$ follows
the similar law (11) but with $\Delta \Phi
=\Phi(\infty)-\Phi(0)$, where $\Phi(0)$ is the gravitational
potential in the center of the Earth. Calculated total life time
is compared with law (11) in Figure~\ref{timelife2}. As is seen
from the comparison total time life exactly follows Boltzmann
law. Fitted temperature of the captured DM in the Earth is equal
to $4900$ K which is about 75\% of the maximal temperature in the
Earth core. Total time life of heavy DM particles in the
gravitational potential of the Earth exceeds the age of the
Earth. This time determines the time of DM accumulation by the
Earth. The accumulation time is equal to the total time life of
the DM particle if it doesn't exceed the age of the Earth, or
simply equals the age of the Earth if total time life is higher
than the Earth age.

\section{DM density on the Earth surface}
\begin{figure}[!hbp]
\centerline{ \epsfxsize=15cm\epsfbox{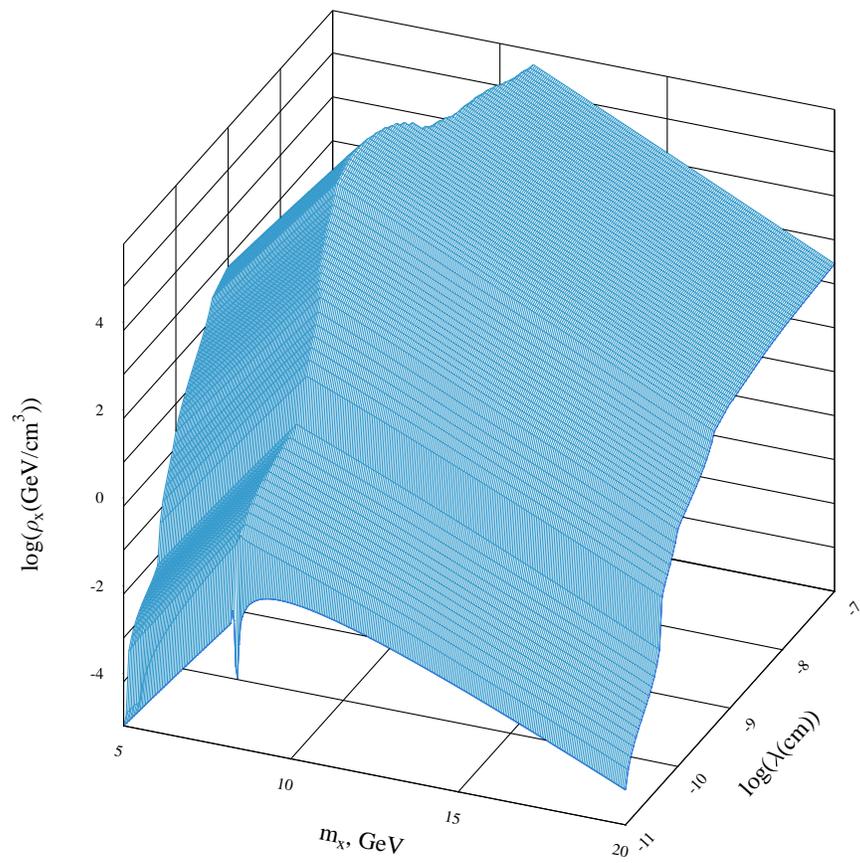}}
\caption{Additional DM density on the Earth surface} \label{DMden}
\end{figure}

Captured by the Earth gravitation DM particles constantly fly out
 of the Earth, stay some time at elliptic orbit and then return to
 the Earth. As the age of the Earth is rather large (about 4.5
 billion years) and the rate with which LRF DM is captured by the Earth
 is also large the amount of DM accumulated in the Earth may be up to
 $10^{37}$ particles. Accumulated by the Earth DM particles can
 produce additional DM density at the Earth surface. This density can
 be calculated as \beq \rho_x=m_x \frac{C\, t_{accum}}{(t_{Earth}+t_{orb}) S_E \bar v_{orb}},
 \eeq
 where $S_E$ is the area of the Earth surface, $t_{orb}$ is the
 average time the DM particles stay at the orbit outside the
 Earth which is usually not larger than 0.1 day, and $\bar v$ is the mean
 DM particle velocity when it leaves the Earth surface.
 In Eq. (12) $C$ is the capture rate of DM particles by the Earth, and $t_{accum}$
 is the accumulation time of DM by the Earth, which is equal to total time life of DM particle captured by the Earth gravitation if it doesn't exceed the Earth age.
 The results of the
 calculations of additional DM density for different interaction length $\lambda$ and DM
 particle mass $m_x$ are presented in Figure~\ref{DMden}. As is
 seen from this figure for masses of DM particles smaller than 15
 GeV the additional DM density could approach rather high values
 higher than $10^6$ times exceeding the mean DM density in our
 Galaxy.

\section{Discussion}
\begin{figure}[!hbp]
\centerline{ \epsfxsize=15cm\epsfbox{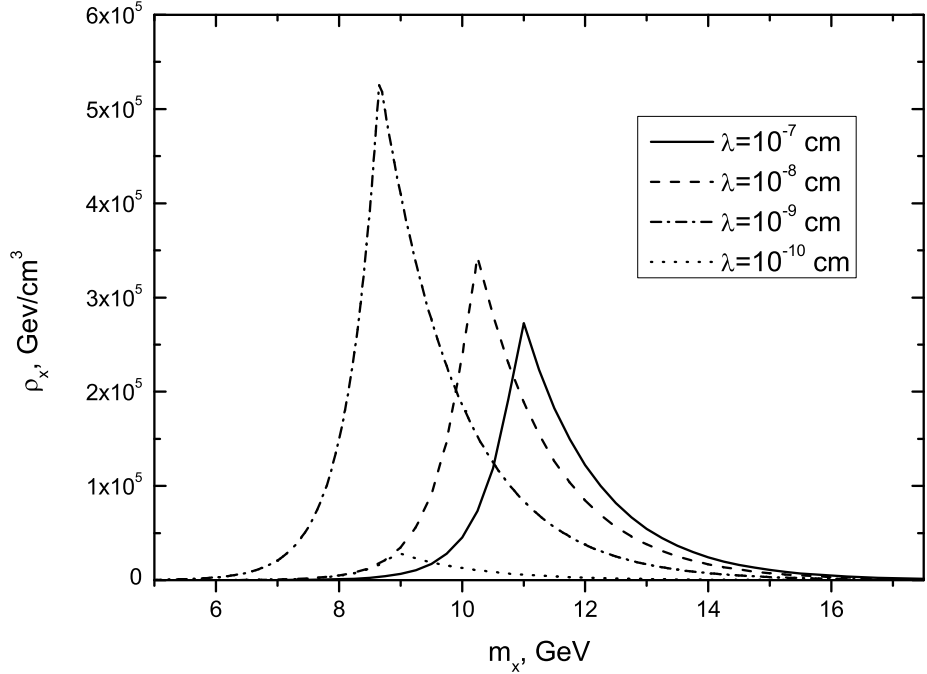}}
\caption{Additional DM density on the Earth surface} \label{DMmx}
\end{figure}

Additional DM density at the Earth surface presented in
Figure~\ref{DMden} has a form of a mountain ridge. Mass
dependence on $\rho_x$ for different $\lambda$ is presented in
Figure~\ref{DMmx}. It has a sharp peak with exponential growth
before peak and exponential falling after the peak. Such a behavior
is determined by twp times in Eq.(12) - $t_{accum}$ and
$t_{Earth}$. These both times rise exponentially with $m_x$ before
$t_{accum}$ reaches saturation determined by the age of the Earth.
After these times ratio of these two times changes from exponential
growth to exponential falling. Capture rate $C$ has a smooth
behavior with $m_x$ and doesn't significantly influence the
$m_x$ behavior (see Fig.~\ref{Caprate2}). The height of the peak
in Fig.~\ref{DMden} only slightly decreases with $\lambda$ so that
it is reasonable to suppose that for $\lambda$ up to $10^{-5}$ cm
there will be a significant accumulation of DM at the Earth surface
due to the considered mechanism. It is to be mentioned that the width
of the peaks is not large (about several GeV at half width) that
is why this mechanism of DM accumulation is selective for definite
masses of DM if the range of LRF is fixed. Also the accumulated
density of DM near the Earth is determined by the value of the
constant $\alpha$ in Eq.(1). For smaller values of $\alpha$ the
density of accumulated DM at the Earth surface will decrease.

\begin{figure}[!hbp]
\centerline{ \epsfxsize=15cm\epsfbox{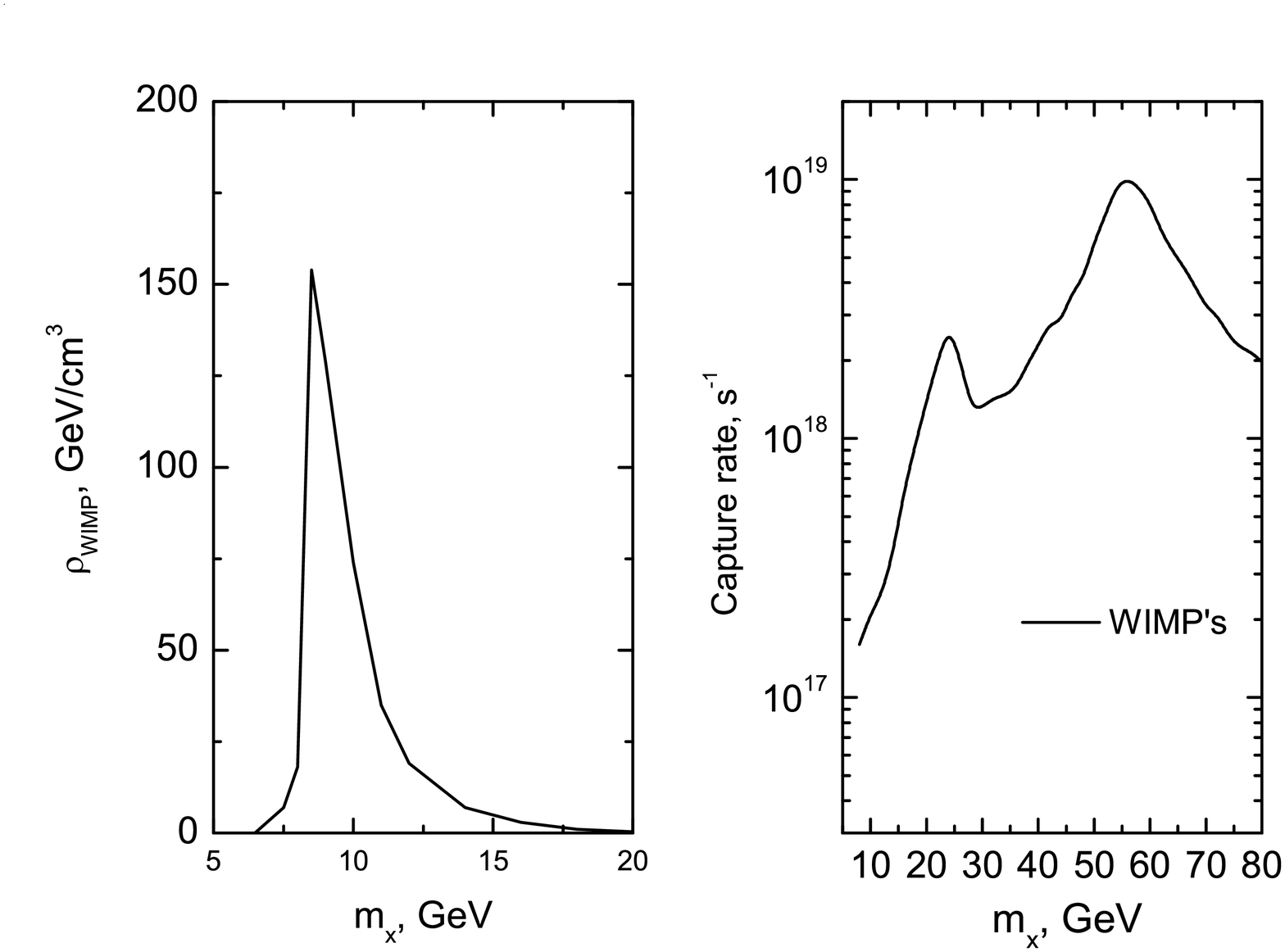}}
\caption{Additional WIMP's density at the Earth surface (left panel)
and capture rate of WIMP's by the Earth (right panel)}
\label{Wimps}
\end{figure}

Additional DM density at the Earth surface for the case of LRF can
be compared with the same density but for the case of WIMP's. Both
in~\cite{Gould} and in a WIMP's is considered the case of heavy Dirac
neutrino. The total scattering cross section on the Earth element
is assumed to be~\cite{Gould}

\beq \sigma=\frac{\mu}{\mu_+^2}Q^2\frac{m m_x}{(\textrm{
GeV})^2}\times 5.2 \times 10^{-40}\,\, \textrm{cm}^2, \eeq where
\beq Q=N-(1-4\sin ^2 \theta_W)Z \approx N-(0.124)Z \eeq and $N,Z$
is the neutron and proton numbers of the Earth element. The calculated Capture
rate using this cross section is presented in
Fig.~\ref{Wimps}. It has two peaks that appear when the WIMP's mass
coincides with that of the Earth element in our model of the Earth.
This capture rate agrees well with that calculated
theoretically in~\cite{Gould}. Additional density of WIMP's at the
Earth surface (due to WIMP's evaporation) has also been calculated. It
is presented in Fig.~\ref{Wimps}. Additional density of WIMP's
achieves its maximal value for the WIMP's mass about 8.5 GeV when
the time life of WIMP's reaches the age of the Earth. It is equal to
150 GeV/cm$^3$ which is about three orders of magnitude smaller
than maximal additional density of DM with LRF. This discrepancy
is due to the fact that the time life in the Earth is much larger
for WIMP's than for long range interaction DM. This is because the
scattering cross section of WIMP's is small so they need large
time before they get in the collision at the energy sufficient  to
escape the Earth. As is seen from Fig.~\ref{timeE} this time increases
for DM with LRF for small $\lambda$ where the cross
section falls as is seen from Fig.~\ref{CS}.

In this paper we have considered only the case of DM capture and
accumulation by the Earth. Besides this direct process there can also
be a two step process. At the first step DM is captured by the Sun
and then due to perturbations additional density of DM near the
Earth is produced~\cite{Serebrov, Lundberg, Damour, Press,
Gould2}. Besides that due to this process DM is accumulated near
the Earth, the velocity spectrum of DM particles is much softer in
such a way that the probability of its capturing by the Earth is
higher in this case than for the case of Galaxy DM. At the second
step DM accumulated at the Earth orbit is captured by the Earth.
Unfortunately this mechanism of accumulation is much more
complicated than the one considered in this paper and there are
also some uncertainties in time life of DM in the Sun system at
the Earth orbit~\cite{Gould3}. On the other hand, this mechanism is
rather interesting because heavy DM particles can be accumulated
by it.

This work was supported by the Russian Foundation for Basic
Research (project nos. 10-02-00217-à, 10-02-00224-à,
11-02-01435-a, 11-02-91000-ANFa) and by the Federal Agency of
Education of the Russian Federation (contract nos. P2427, P2500,
P2540), by the Ministry of Education and Science of the Russian
Federation (contract nos. 02.740.11.0532, 14.740.11.0083).

\end{document}